\documentclass[journal=jacsat,manuscript=article]{achemso}

\usepackage[version=3]{mhchem} % Formula subscripts using \ce{}
\usepackage{amsmath}
\usepackage{braket}
\usepackage{xcolor}
\usepackage{amsthm,amsmath,amssymb}
\usepackage{physics}
\usepackage{bm}
\usepackage{accents}
\newlength{\dhatheight}

\author{Yu Wang}
\affiliation{Department of Chemistry, School of Science, Westlake University, Hangzhou 310024 Zhejiang, China}
\altaffiliation
{Institute of Natural Sciences, Westlake Institute for Advanced Study, Hangzhou 310024 Zhejiang, China}

\author{Ruihao Bi}
\affiliation{Department of Chemistry, School of Science, Westlake University, Hangzhou 310024 Zhejiang, China}
\altaffiliation
{Institute of Natural Sciences, Westlake Institute for Advanced Study, Hangzhou 310024 Zhejiang, China}

\author{Wei Liu}
\affiliation{Department of Chemistry, School of Science, Westlake University, Hangzhou 310024 Zhejiang, China}
\altaffiliation
{Institute of Natural Sciences, Westlake Institute for Advanced Study, Hangzhou 310024 Zhejiang, China}

\author{Jiayue Han}
\affiliation{Department of Chemistry, School of Science, Westlake University, Hangzhou 310024 Zhejiang, China}
\altaffiliation
{Institute of Natural Sciences, Westlake Institute for Advanced Study, Hangzhou 310024 Zhejiang, China}

\author{Wenjie Dou}
\email{douwenjie@westlake.edu.cn}
\affiliation{Department of Chemistry, School of Science, Westlake University, Hangzhou 310024 Zhejiang, China}
\altaffiliation
{Institute of Natural Sciences, Westlake Institute for Advanced Study, Hangzhou 310024 Zhejiang, China}

\title
  {Mixed Quantum-Classical Approaches to Spin Current and Polarization Dynamics in Chiral Molecular Junctions}

\begin{document}

%%%%%%%%%%%%%%%%%%%%%%%%%%%%%%%%%%%%%%%%%%%%%%%%%%%%%%%%%%%%%%%%%%%%%
%% The "tocentry" environment can be used to create an entry for the
%% graphical table of contents. It is given here as some journals
%% require that it is printed as part of the abstract page. It will
%% be automatically moved as appropriate.
%%%%%%%%%%%%%%%%%%%%%%%%%%%%%%%%%%%%%%%%%%%%%%%%%%%%%%%%%%%%%%%%%%%%%

%\begin{tocentry}

%Some journals require a graphical entry for the Table of Contents.
%This should be laid out ``print ready'' so that the sizing of the
%text is correct.

%Inside the \texttt{tocentry} environment, the font used is Helvetica
%8\,pt, as required by \emph{Journal of the American Chemical
%Society}.

%The surrounding frame is 9\,cm by 3.5\,cm, which is the maximum
%permitted for  \emph{Journal of the American Chemical Society}
%graphical table of content entries. The box will not resize if the
%content is too big: instead it will overflow the edge of the box.

%This box and the associated title will always be printed on a
%separate page at the end of the document.

%\end{tocentry}

%%%%%%%%%%%%%%%%%%%%%%%%%%%%%%%%%%%%%%%%%%%%%%%%%%%%%%%%%%%%%%%%%%%%%
%% The abstract environment will automatically gobble the contents
%% if an abstract is not used by the target journal.
%%%%%%%%%%%%%%%%%%%%%%%%%%%%%%%%%%%%%%%%%%%%%%%%%%%%%%%%%%%%%%%%%%%%%
\begin{abstract}
Chiral molecular junctions offer a promising platform for realizing chiral-induced spin selectivity (CISS), where spin filtering occurs without external magnetic fields. Here, we investigate spin transport in such junctions by combining quantum master equation (QME) methods for purely electronic dynamics with surface hopping (SH) and mean-field Ehrenfest (MF) approaches to incorporate electron–phonon coupling. Our results show that transient spin polarization arises but ultimately decays to zero at long times. We find that bias voltage, molecular length, and spin–orbit coupling (SOC) strongly influence the spin current dynamics: higher bias enhances spin current but reduces polarization, while longer molecules and stronger SOC amplify transient polarization. Including electron–phonon coupling modifies current–voltage characteristics, enhancing spin currents at intermediate bias but suppressing them at high bias, while leaving the polarization dynamics largely unchanged. These findings highlight the interplay between electronic and vibrational effects in CISS and provide guidance for designing molecular spintronic devices.

\end{abstract}

\section{INTRODUCTION}

The interplay between molecular chirality and spin-dependent electron transport has garnered significant interest due to the discovery of the Chiral-Induced Spin Selectivity (CISS) effect\cite{bloom2024chiral}. This phenomenon, wherein chiral molecules preferentially transmit electrons of a particular spin orientation, challenges conventional understanding of spin polarization, especially in systems composed of light, nonmagnetic elements\cite{naaman2019chiral}. CISS offers a promising route toward molecular spintronics, enabling spin filtering, spin injection, and even magnetoresistance control without relying on magnetic materials or external magnetic fields\cite{ben2017magnetization,naaman2020chiral,firouzeh2024chirality}.

Experimental observations of CISS span a diverse range of chiral systems, including helical peptides\cite{abendroth2019spin,torres2020reinforced, theiler2023detection}, DNA strands\cite{gohler2011spin,alam2015spin,das2025spin}, and synthetic helicenes\cite{kiran2016helicenes,qian2022chiral, sun2024inverse}, revealing robust spin polarization effects under both electrical and optical excitations.
At the same time, recent experiments on single-molecule junctions have reported no detectable spin polarization, indicating that in the coherent transport regime the probability of a spin-flip transition is less than $10^{-6}$, rendering the effect too small to be experimentally observed\cite{li2025too}.
These seemingly contrasting results highlight the complexity of CISS and the influence of experimental conditions, such as molecular environment, ensemble effects, and coupling to electrodes.
However, the theoretical foundation of CISS remains under active debate. Key challenges include explaining the emergence of significant spin polarization from systems with weak intrinsic spin–orbit coupling (SOC), capturing dynamical effects during electron transport, and understanding the role of electronic coherence and decoherence in molecular junctions\cite{evers2022theory}.

Existing theoretical approaches—ranging from static scattering theory\cite{yeganeh2009chiral,eremko2013spin,ghazaryan2020analytic} and non-equilibrium Green’s function (NEGF) methods\cite{guo2012spin,naskar2023chiral,garcia2023nonequilibrium,fransson2025chiral,liu2025enhancement} to time-dependent wavepacket dynamics\cite{caetano2016spin}—have provided valuable insights but often yield inconsistent predictions depending on assumptions about electronic structure, SOC strength, and environmental interactions. Notably, certain models reproduce pronounced spin selectivity, while others, even when applied to similar systems, fail to predict any spin polarization at all\cite{nurenberg2019evaluation}.

In this work, we combine quantum master equation (QME) method for purely electronic dynamics with surface hopping (SH) and mean-field Ehrenfest (MF) approaches to incorporate electron–phonon coupling, to simulate spin transport for chiral molecule junctions.
This framework allows us to capture both coherent spin transport and vibrationally assisted processes in chiral molecular junctions. 
We adopt the chiral molecular junction model introduced in Ref. \citenum{smorka2025influence}, which builds on Ref. \citenum{fransson2020vibrational}, with the only modification being the replacement of quantum nuclear vibrations by classical ones.
We systematically investigate how bias voltage, molecular length, SOC strength, and electron–phonon interactions affect the time-dependent evolution of spin current and spin polarization.
Our results show that the spin current increases with stronger bias. Spin polarization appears transiently on the sub-picosecond timescale but always decays to zero at long times. Stronger SOC and longer molecular backbones enhance the magnitude of the transient polarization. Electron–phonon coupling further modifies current–voltage characteristics, enhancing currents at intermediate bias but suppressing them at high bias, while leaving the polarization dynamics largely unaffected.
These findings provide new insights into the interplay between SOC and vibrational effects in chiral systems. They highlight the critical role of transient dynamics in understanding spin selectivity and suggest strategies for optimizing spin transport in molecular devices. Our work paves the way for designing chiral molecular junctions with tailored spintronic functionalities and for exploring vibrationally assisted spin control in nanoscale systems.

\section{RESULTS AND DISCUSSIONS}

Following Refs. \citenum{smorka2025influence} and \citenum{fransson2020vibrational}, we consider a chiral molecule as the system, modeled as a helical structure with radius $a$ and pitch $c$.
The spatial coordinates of site $m$ are given by
\begin{gather}
    \bm{r}_m=\left(a\cos\phi_m, a\sin\phi_m, (m-1)c/(N-1)\right), \\
    \phi_m = 2\pi(m-1)N_{\rm laps}/(N-1),
\end{gather}
where $N=N_{\rm laps}\times N_{\rm ions}$ is the total number of sites, with $N_{\rm laps}$ denoting the number of helical turns and $N_{\rm ions}$ the number of sites per turn.

The system Hamiltonian follows the chiral molecular model:
\begin{gather}\label{H_chiral}
    \hat{H}_{s} = \sum_{m=1}^N \epsilon_0\hat{d}_m^{\dagger}\hat{d}_m+\frac{1}{2}\hbar\omega_0(\bm{R}^2+\bm{P}^2)-\sum_{m=1}^{N-1}\left(\hat{d}_m^{\dagger}\hat{d}_{m+1}+H.c.\right)\left[t_0+\sqrt{2}t_1\bm{R}_m \right] \\ \notag
    + \sum_{m=1}^{N-2}i\left(\hat{d}_m^{\dagger}\bm{\nu}_m^+\cdot\bm{\sigma}\hat{d}_{m+2}\right)\left[\lambda_0+\sqrt{2}\lambda_1\bm{R}_m \right]
\end{gather}
The molecular Hamiltonian describes a chiral tight-binding chain of $N$ sites coupled to a single vibrational mode of frequency $\omega_0$, incorporating both nearest-neighbor electron–phonon interactions and next-nearest-neighbor spin–orbit coupling (SOC).
The first term accounts for the on-site electronic energy $\epsilon_0$ at each site, while the second term represents the energy of the classical vibrational mode described by nuclear position $\bm{R}$ and momentum $\bm{P}$. The third term captures electron hopping between adjacent sites with amplitude $t_0$,  which is modulated linearly by the phonon displacement with coupling constant $t_1$,  thus enabling phonon-assisted charge transfer. The last term introduces an intrinsic SOC between next-nearest neighbors, proportional to $i\bm{\nu}_m^s\cdot\bm{\sigma}$, where the unit vector $\bm{\nu}_m^s=\bm{d}_{m+s}\times \bm{d}_{m+2s}$ ($s=\pm 1$) encodes the local chirality, with $\bm{d}_{m+s}=(\bm{r}_m-\bm{r}_{m+s})/|\bm{r}_m-\bm{r}_{m+s}|$. This geometric factor ensures that the SOC strength reflects the handedness of the helical backbone.
This SOC is characterized by a bare strength $\lambda_0$ and a phonon-modulated contribution $\lambda_1$, allowing the vibrational mode to dynamically influence spin-dependent transport. 
Such SOC amplitude is also modulated by the linear phonon displacement.

Additionally, we include the fermionic bath, representing the metal electrodes, along with the system–bath coupling in the total Hamiltonian, which is presented in the Theoretical Framework section.

The parameters used in this paper: $\epsilon_0=-240$ meV, $\omega_0=0.4$ meV, $t_0=40$ meV, $t_1=4$ meV, $\lambda_0=1$ meV (10 meV), $\lambda_1=0.1$ meV, $k_BT=25$ meV.
For system-bath coupling $\Gamma_L^A=\Gamma_L(\mathcal{I}+A\sigma_z)$, $\Gamma_L=\Gamma_R=10$ meV, $A=+(-)\frac{1}{2}$ for spin up (spin down) current. $\mu_L=-\mu_R$, bias voltage $\Phi=(\mu_L-\mu_R)/e$.

\subsection{WITHOUT NUCLEAR MOTION}

We first investigate the spin polarization in the absence of nuclear motion, focusing on the effects of the molecular length and spin–orbit coupling (SOC) strength. The current is calculated using a single-particle quantum master equation (QME) approach.

\begin{figure}
    \centering
    \includegraphics[width=0.9\linewidth]{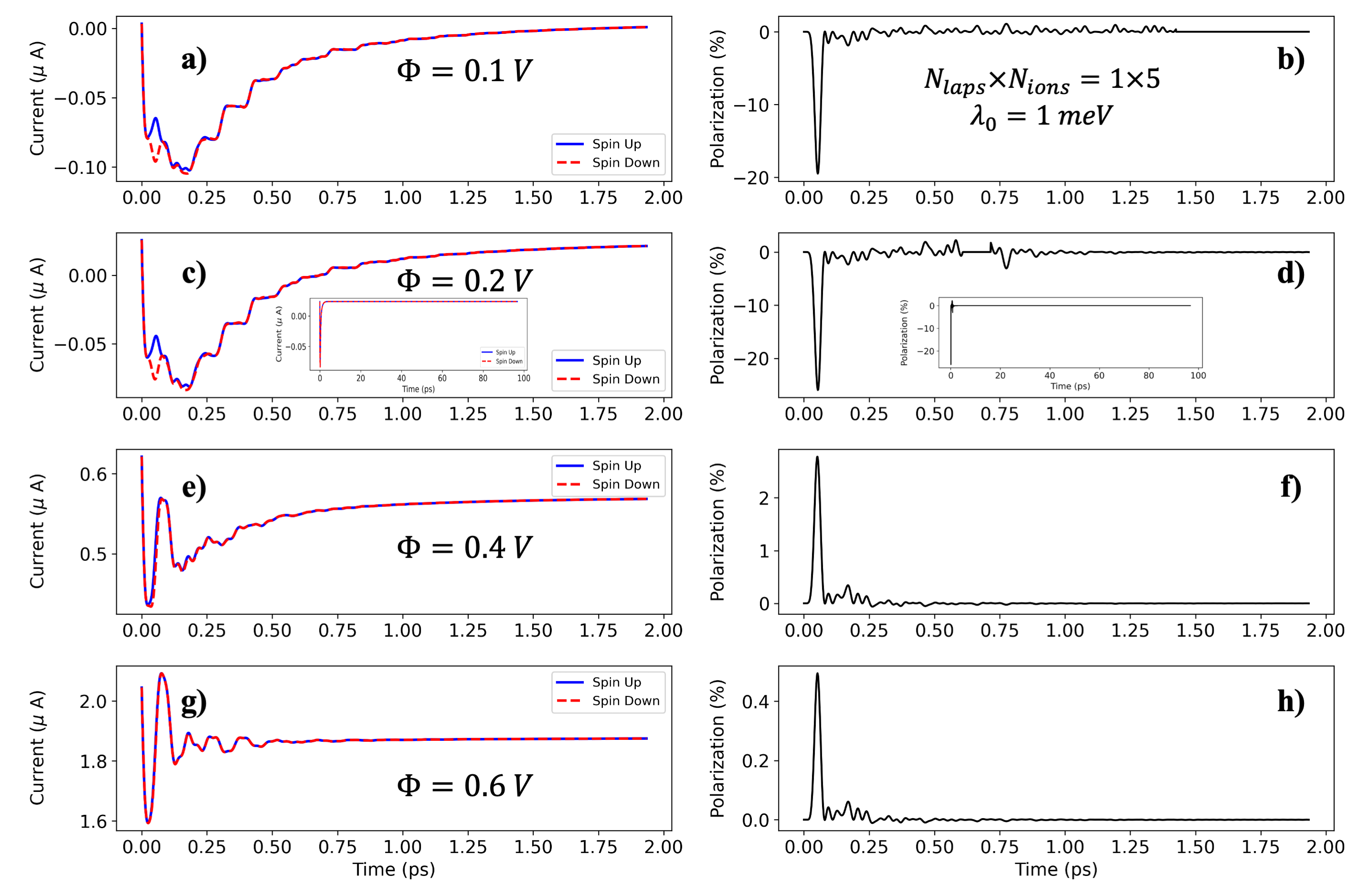}
    \caption{Time-dependent spin current (left panels) and spin polarization (right panels) for a $N_{\rm laps}\times N_{\rm ions}=1\times5$ chiral molecule under different biases: (a, b) $\Phi=0.1$ V, (c, d) $\Phi=0.2$ V, (e, f) $\Phi=0.4$ V, and (g, h) $\Phi=0.6$ V. Here, we applied a small SOC strength where $\lambda_0=1$ meV.}
    \label{fig:1}
\end{figure}

Figure $\ref{fig:1}$ shows the time-dependent current (left panels) and spin polarization (right panels) for the chiral tight-binding system under different applied voltage biases $\Phi$. Here, we consider a short chiral molecule with $N_{\rm laps}\times N_{\rm ions}=1\times 5$. Left panels (a, c, e, g) display the spin-resolved currents for spin-up (blue solid line) and spin-down (red dashed line). Right panels (b, d, f, h) show the corresponding spin polarization, defined as $\frac{I_{\uparrow}-I_{\downarrow}}{I_{\uparrow}+I_{\downarrow}}\times 100\%$.
In the low-bias regime ($\Phi = 0.1$ V and $0.2$ V), the currents are very small, initially negative due to the transient response, and nearly identical for spin-up and spin-down on the picosecond timescale. The spin polarization exhibits large short-time fluctuations (up to $\sim 20\%$) but decays to nearly zero in the long-time steady state. The inset in Fig.~\ref{fig:1}b and c shows the current and polarization at 100 ps, where the current gradually stabilizes after the initial transient, and the spin polarization, despite strong early fluctuations, ultimately vanishes.
At intermediate bias ($\Phi=0.4$ V), the current increases to the sub-microampere range. The spin polarization shows a sharp transient peak (about $2\%$), yet it also vanishes in the steady state.
In the high-bias regime ($\Phi=0.6$ V), the current reaches the microampere scale ($\sim 1.8$–$2.0\ \mu$A). The spin polarization is strongly suppressed, showing only a small transient peak ($\sim 0.5\%$) before decaying to zero.

\begin{figure}
    \centering
    \includegraphics[width=0.9\linewidth]{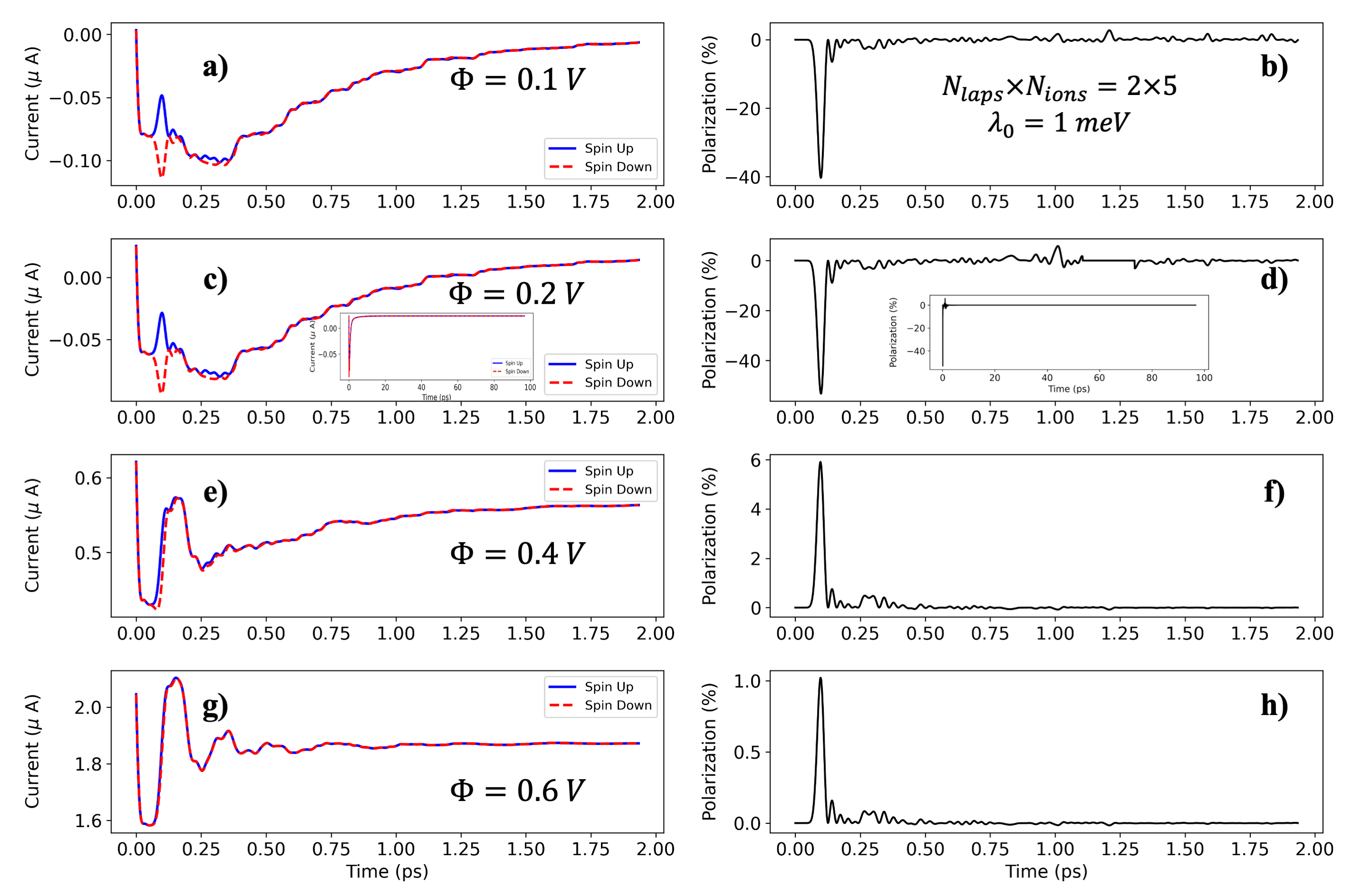}
    \caption{Time-dependent spin current (left panels) and spin polarization (right panels) for a $N_{\rm laps}\times N_{\rm ions}=2\times5$ chiral molecule under different biases: (a, b) $\Phi=0.1$ V, (c, d) $\Phi=0.2$ V, (e, f) $\Phi=0.4$ V, and (g, h) $\Phi=0.6$ V. Here, we applied a small SOC strength where $\lambda_0=1$ meV.}
    \label{fig:2}
\end{figure}

Figure $\ref{fig:2}$ presents the corresponding results for a longer chiral molecule with $N_{\rm laps}\times N_{\rm ions}=2\times 5$. Similar to Fig.~\ref{fig:1}, the left panels present the spin-resolved currents for spin-up (blue solid line) and spin-down (red dashed line), while the right panels display the corresponding spin polarization.
Compared to the shorter molecule, the overall current amplitude changes only slightly across all bias regimes, indicating that current transport is not strongly affected by the increased length on the timescale considered. In contrast, the transient spin polarization is significantly enhanced.
At low bias ($\Phi=0.1$ V), the polarization exhibits large short-time fluctuations reaching nearly $40\%$. For $\Phi=0.2$ V, the transient polarization exceeds $40\%$. The long-time results (100 ps) shown in the inset of Fig.~\ref{fig:2}b and c indicate that the polarization decays to nearly zero in the steady state.
At intermediate bias ($\Phi=0.4$ V), the transient polarization is reduced but still reaches about $6\%$, while at high bias ($\Phi=0.6$ V) it remains finite at approximately $1\%$. In all cases, however, the polarization decays to nearly zero in the steady state.

\begin{figure}
    \centering
    \includegraphics[width=0.9\linewidth]{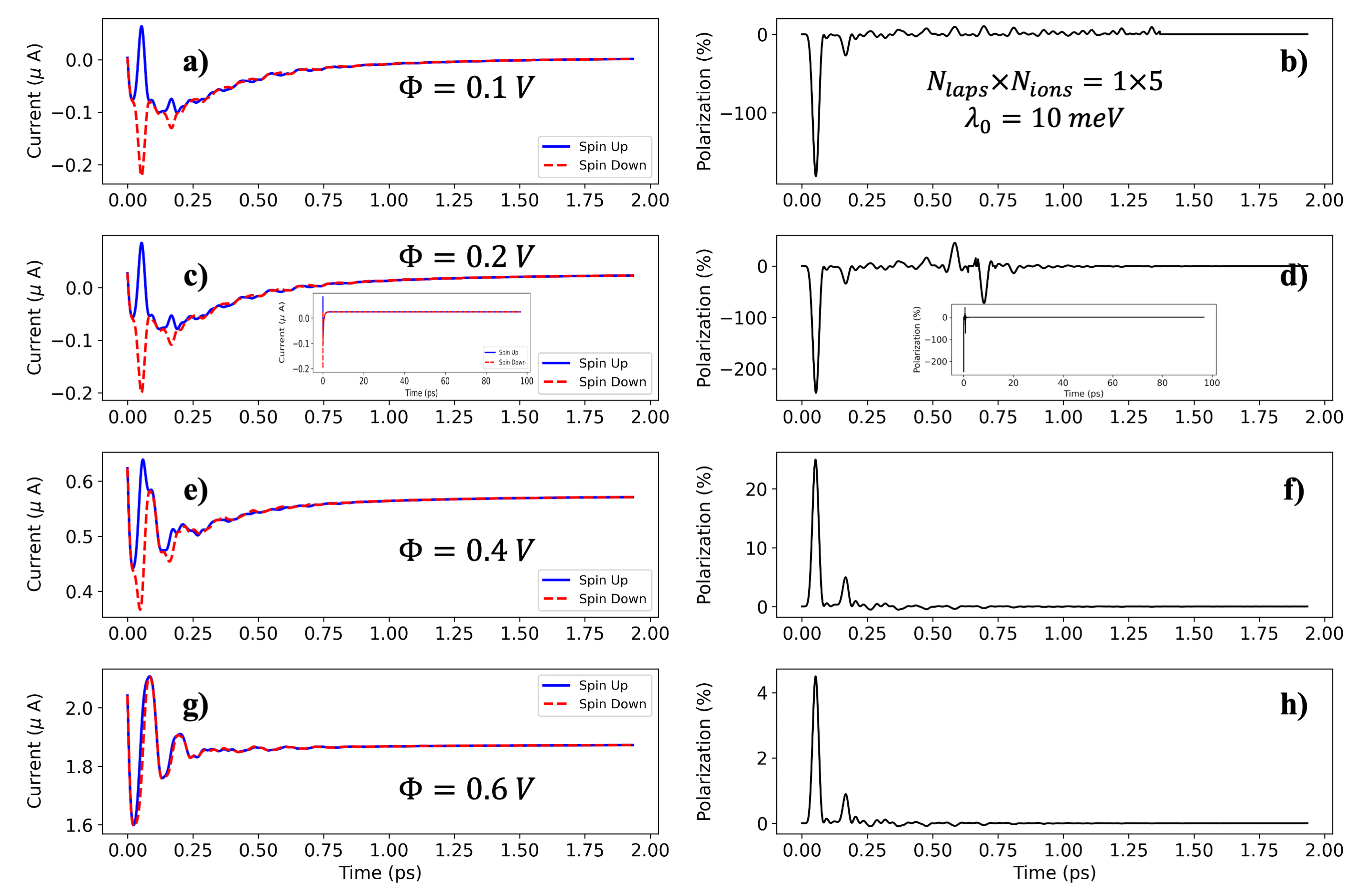}
    \caption{Time-dependent spin current (left panels) and spin polarization (right panels) for a $N_{\rm laps}\times N_{\rm ions}=1\times5$ chiral molecule under different biases: (a, b) $\Phi=0.1$ V, (c, d) $\Phi=0.2$ V, (e, f) $\Phi=0.4$ V, and (g, h) $\Phi=0.6$ V. Here, we applied a large SOC strength where $\lambda_0=10$ meV.}
    \label{fig:3}
\end{figure}

Figure $\ref{fig:3}$ shows the short chiral molecule under different bias voltages $\Phi$ with the SOC strength increased by an order of magnitude compared to Fig.~\ref{fig:1}.
The enhancement of SOC leads to a dramatic increase in the transient spin polarization. At low bias ($\Phi=0.1$ V), the polarization exceeds $100\%$, arising from the fact that the total current ($I_{\uparrow}+I_{\downarrow}$) approaches zero near the transient crossing point, which amplifies the polarization ratio.
When the bias is $\Phi=0.2$ V, the transient polarization reaches values above $200\%$. The long-time behavior depicted in the inset of Fig.~\ref{fig:3}b and c demonstrates that the polarization diminishes to nearly zero in the steady state.
At intermediate bias ($\Phi=0.4$ V), the polarization still shows a sizable transient peak of more than $20\%$, while at high bias ($\Phi=0.6$ V) it remains finite at around $4\%$. In all cases, however, the polarization decays to nearly zero at long times.

\begin{figure}
    \centering
    \includegraphics[width=0.9\linewidth]{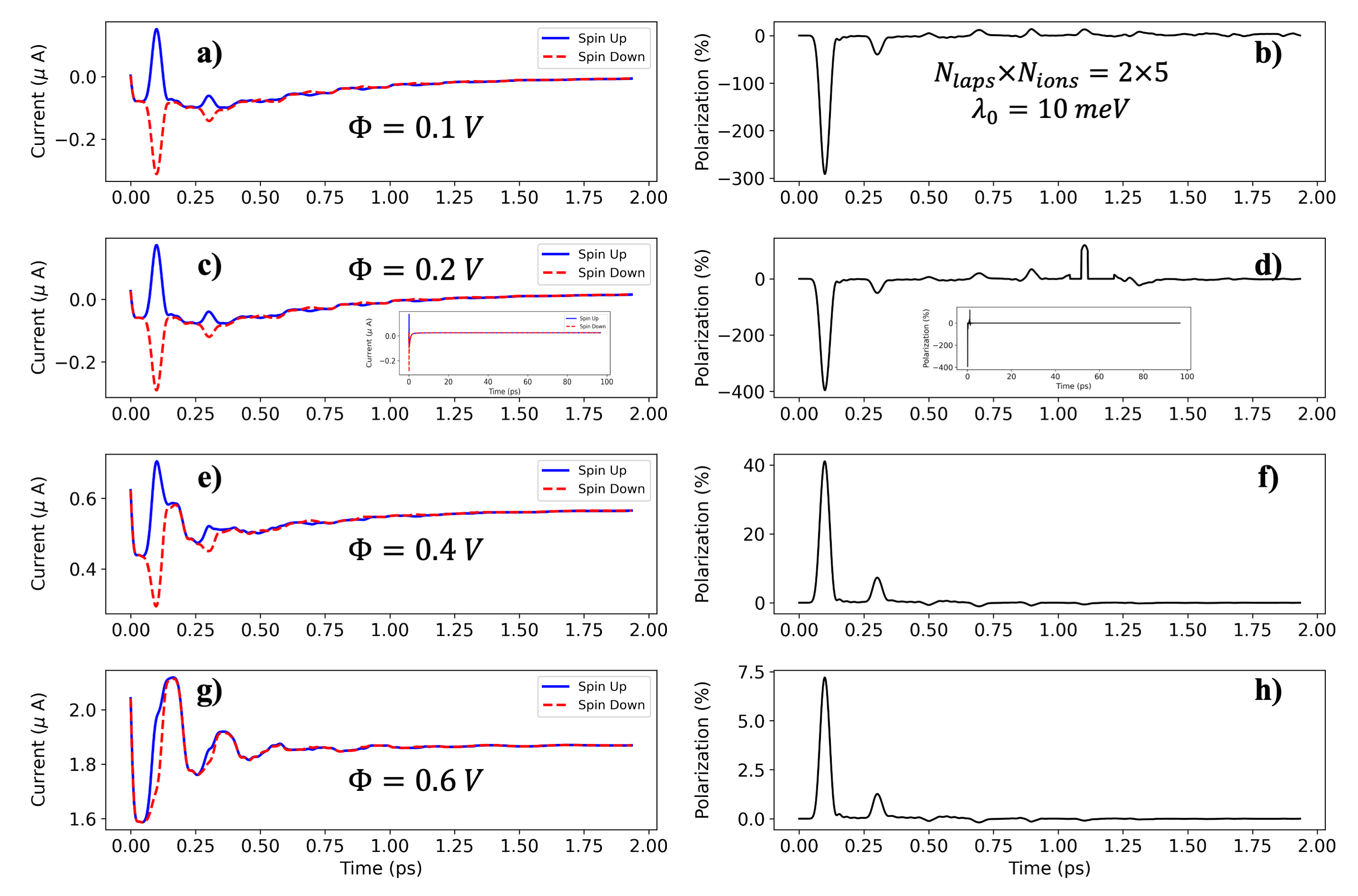}
    \caption{Time-dependent spin current (left panels) and spin polarization (right panels) for a $N_{\rm laps}\times N_{\rm ions}=2\times5$ chiral molecule under different biases: (a, b) $\Phi=0.1$ V, (c, d) $\Phi=0.2$ V, (e, f) $\Phi=0.4$ V, and (g, h) $\Phi=0.6$ V. Here, we applied a large SOC strength where $\lambda_0=10$ meV.}
    \label{fig:4}
\end{figure}

Figure $\ref{fig:4}$ presents the results for the long chiral molecule with the SOC strength enhanced by a factor of ten compared to Fig.~\ref{fig:2}.
With both increased molecular length and stronger SOC, the transient spin polarization is markedly amplified. At low bias ($\Phi=0.1$ V), the polarization reaches nearly $300\%$, while at $\Phi=0.2$ V it approaches $400\%$. The inset of Fig.~\ref{fig:4}b and c shows that, at 100 ps, the polarization has essentially decayed to zero, indicating a vanishing steady-state value.
At intermediate bias ($\Phi=0.4$ V), the polarization remains sizable at over $40\%$, and at high bias ($\Phi=0.6$ V) it still exceeds $7\%$. Despite these large transient enhancements, the polarization decays toward zero in the steady state, consistent with the behavior observed in shorter molecules and weaker SOC cases.

These results indicate that, for the chiral tight-binding model considered here, transient spin polarization can be substantial, but the long-time steady-state polarization vanishes in the absence of nuclear motion.

\subsection{WITH NUCLEAR MOTION}

In this section, we introduce electron–phonon coupling into both the nearest-neighbor (NN) hopping and next-nearest-neighbor (NNN) SOC terms. The spin current dynamics are evaluated using the surface hopping (SH) and mean-field Ehrenfest (MF) methods.

\begin{figure}
    \centering
    \includegraphics[width=0.9\linewidth]{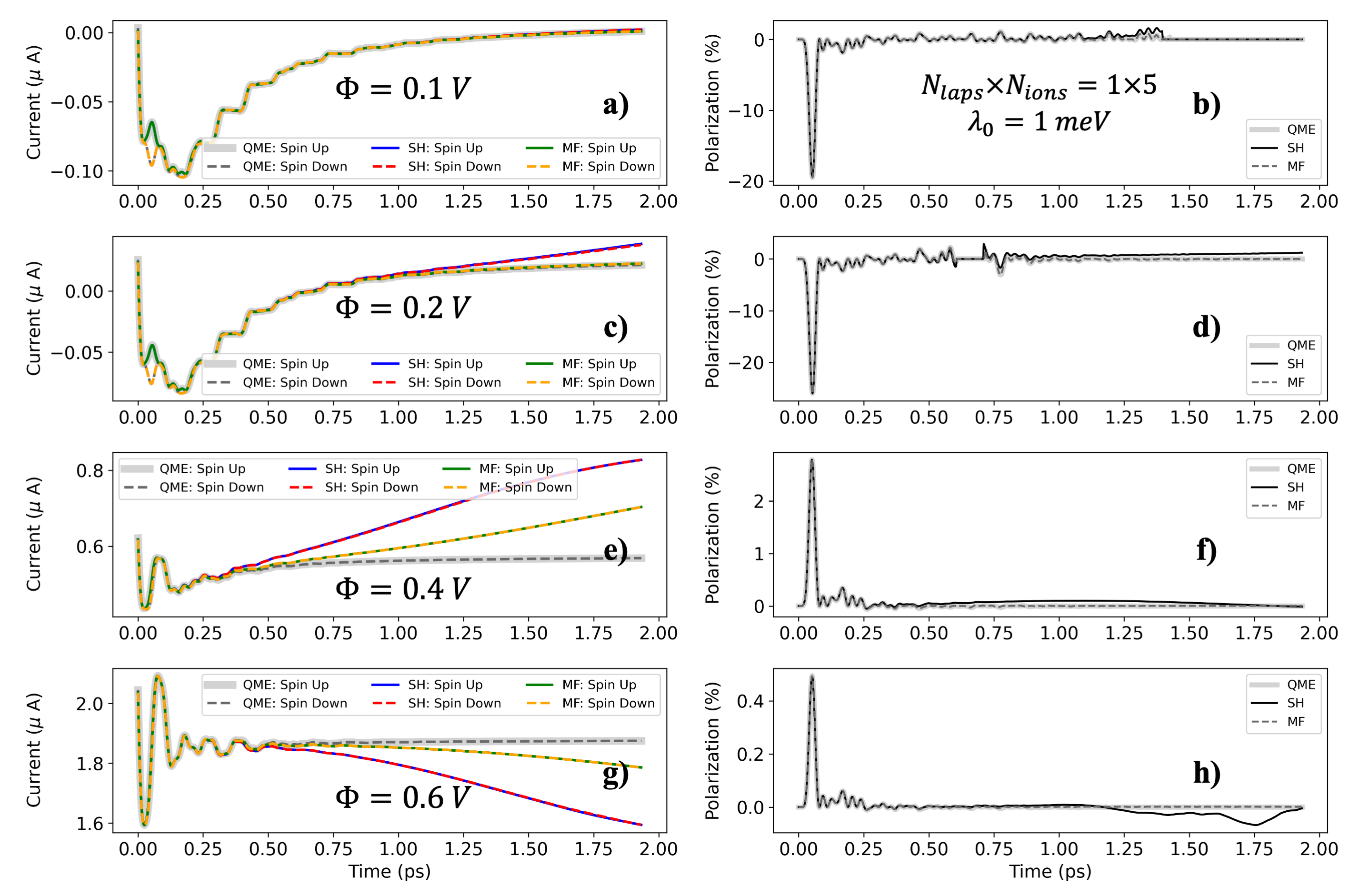}
    \caption{Time-dependent spin current (left panels) and spin polarization (right panels) for a $N_{\rm laps}\times N_{\rm ions}=1\times5$ chiral molecule with electron-phonon couplings under different biases: (a, b) $\Phi=0.1$ V, (c, d) $\Phi=0.2$ V, (e, f) $\Phi=0.4$ V, and (g, h) $\Phi=0.6$ V. Here, we applied a small SOC strength where $\lambda_0=1$ meV.}
    \label{fig:5}
\end{figure}

In Figure $\ref{fig:5}$, the left panels (a, c, e, g) display the time-dependent spin-resolved currents under different bias voltages $\Phi$, while the right panels (b, d, f, h) show the corresponding spin polarizations. Results obtained from the QME (gray lines) represent the reference case without nuclear motion. The SH (blue/red lines) and MF (green/orange lines) dynamics account for electron–phonon coupling.
At low bias ($\Phi=0.1$ V), the transient spin currents obtained from QME, SH, and MF coincide almost perfectly for the first 2 ps, indicating that in such short chiral molecule, the transient electronic transport is dominated by elastic tunneling. In this regime, the available bias window is too small to activate inelastic phonon channels, so nuclear motion does not play a significant role.

When the bias is increased to $\Phi=0.2$ V, SH dynamics predicts a modest current enhancement relative to QME, while MF fails to reproduce this effect. The enhancement can be attributed to phonon-assisted tunneling, where electronic transitions couple to vibrational displacements, effectively opening additional inelastic channels\cite{galperin2007molecular,mitra2004phonon,vdovin2016phonon}. The SH method captures stochastic hopping between adiabatic surfaces, which better describes the phonon-induced broadening of transport channels, whereas MF, being an averaged mean-field approach, underestimates this effect\cite{wang2013charge,chen2025correct}.

At an intermediate bias ($\Phi=0.4$ V), the role of electron–phonon coupling becomes more pronounced. Both SH and MF yield significantly larger steady-state currents compared to QME. In this regime, the bias window is sufficiently large that multiple inelastic channels open, and phonon scattering facilitates electron flow by effectively broadening the transmission spectrum. This behavior can be viewed as phonon-assisted conduction, where nuclear fluctuations reduce localization effects and increase electronic delocalization along the transport pathway\cite{reed2008inelastic, lorente2000theory}.

At high bias ($\Phi=0.6$ V), however, electron–phonon coupling suppresses the current relative to QME. In this regime, strong driving enhances electron–phonon scattering, which introduces decoherence and backscattering, thereby reducing the net current. Since the phonon frequency is very small (0.4 meV) compared to the bias, the lattice behaves almost like a classical fluctuating environment; instead of opening new transport channels, these fluctuations primarily act to randomize electronic motion and dissipate current. This explains why the current is lower than in the purely electronic case at large bias\cite{krause2015thermodynamics,seelig2003electron}.

In contrast to the strong bias-dependent effects on the spin-resolved current, the spin polarization (right panels of Fig. \ref{fig:5}) remains essentially unaffected by nuclear motion across all voltages. Both SH and MF reproduce the QME polarization, which shows short-lived transient peaks but vanishes in the long-time steady state.

\begin{figure}
    \centering
    \includegraphics[width=0.9\linewidth]{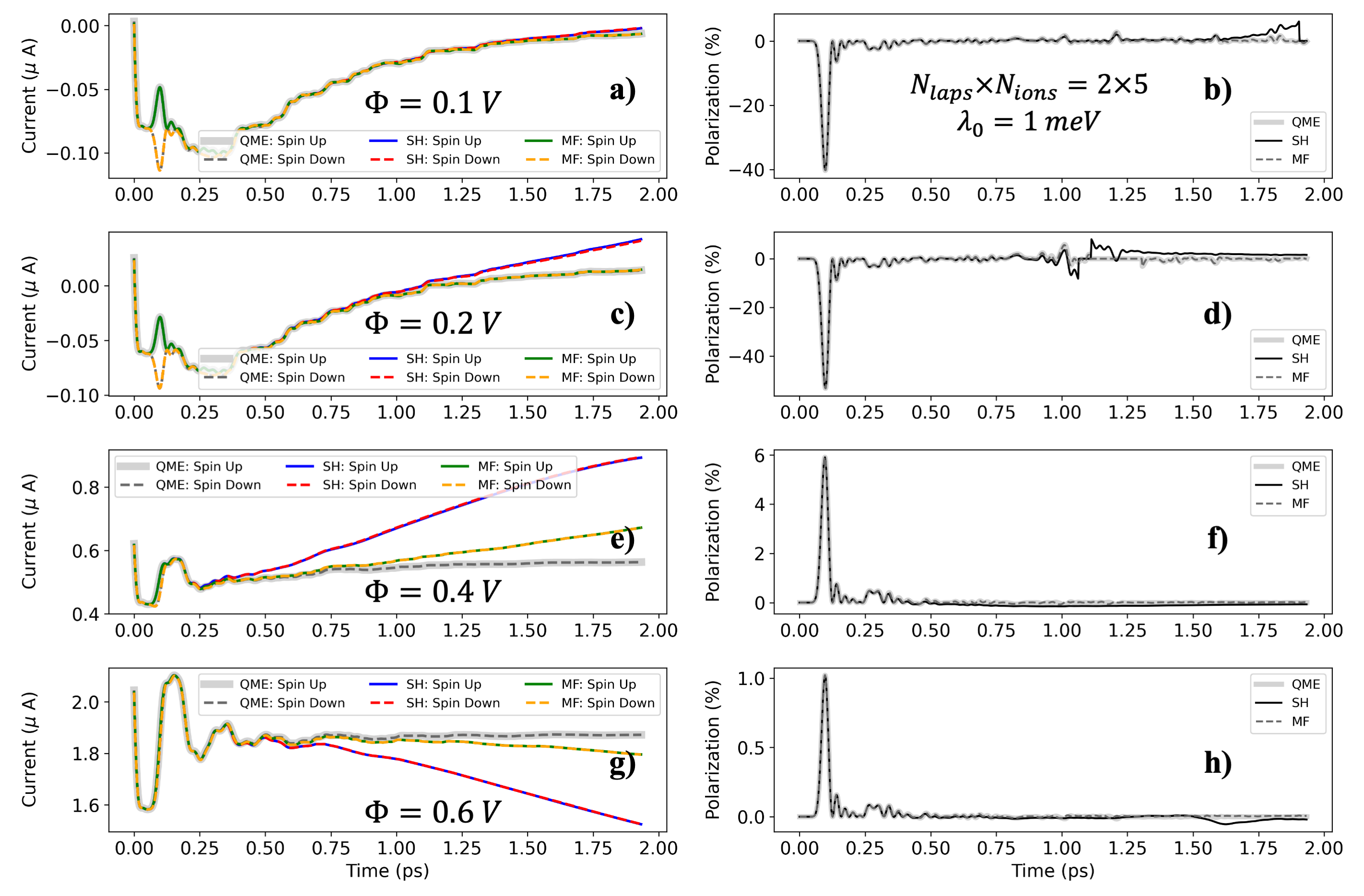}
    \caption{Time-dependent spin current (left panels) and spin polarization (right panels) for a $N_{\rm laps}\times N_{\rm ions}=2\times5$ chiral molecule with electron-phonon couplings under different biases: (a, b) $\Phi=0.1$ V, (c, d) $\Phi=0.2$ V, (e, f) $\Phi=0.4$ V, and (g, h) $\Phi=0.6$ V. Here, we applied a small SOC strength where $\lambda_0=1$ meV.}
    \label{fig:6}
\end{figure}

\begin{figure}
    \centering
    \includegraphics[width=0.9\linewidth]{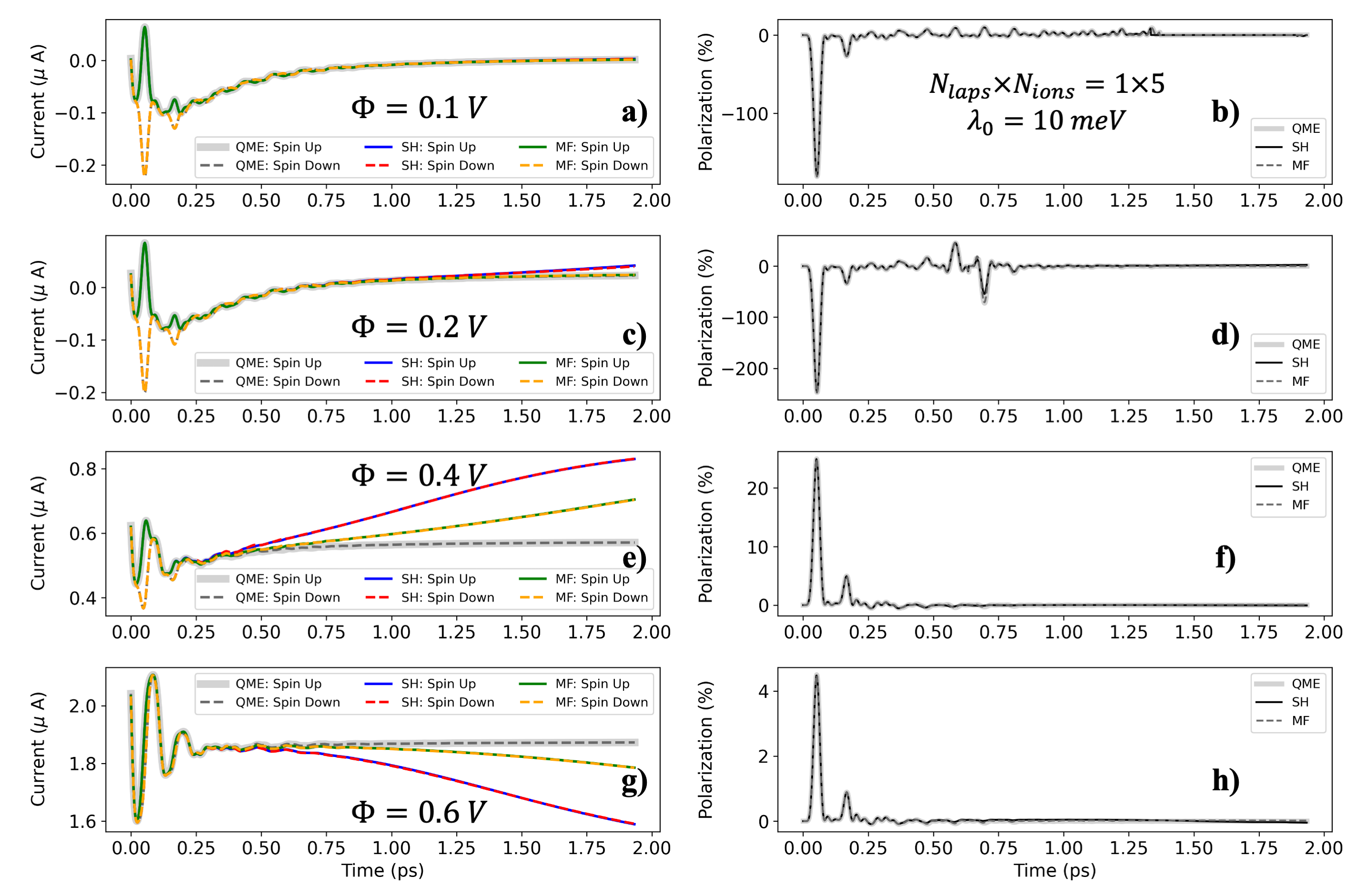}
    \caption{Time-dependent spin current (left panels) and spin polarization (right panels) for a $N_{\rm laps}\times N_{\rm ions}=1\times5$ chiral molecule with electron-phonon couplings under different biases: (a, b) $\Phi=0.1$ V, (c, d) $\Phi=0.2$ V, (e, f) $\Phi=0.4$ V, and (g, h) $\Phi=0.6$ V. Here, we applied a large SOC strength where $\lambda_0=10$ meV.}
    \label{fig:7}
\end{figure}

\begin{figure}
    \centering
    \includegraphics[width=0.9\linewidth]{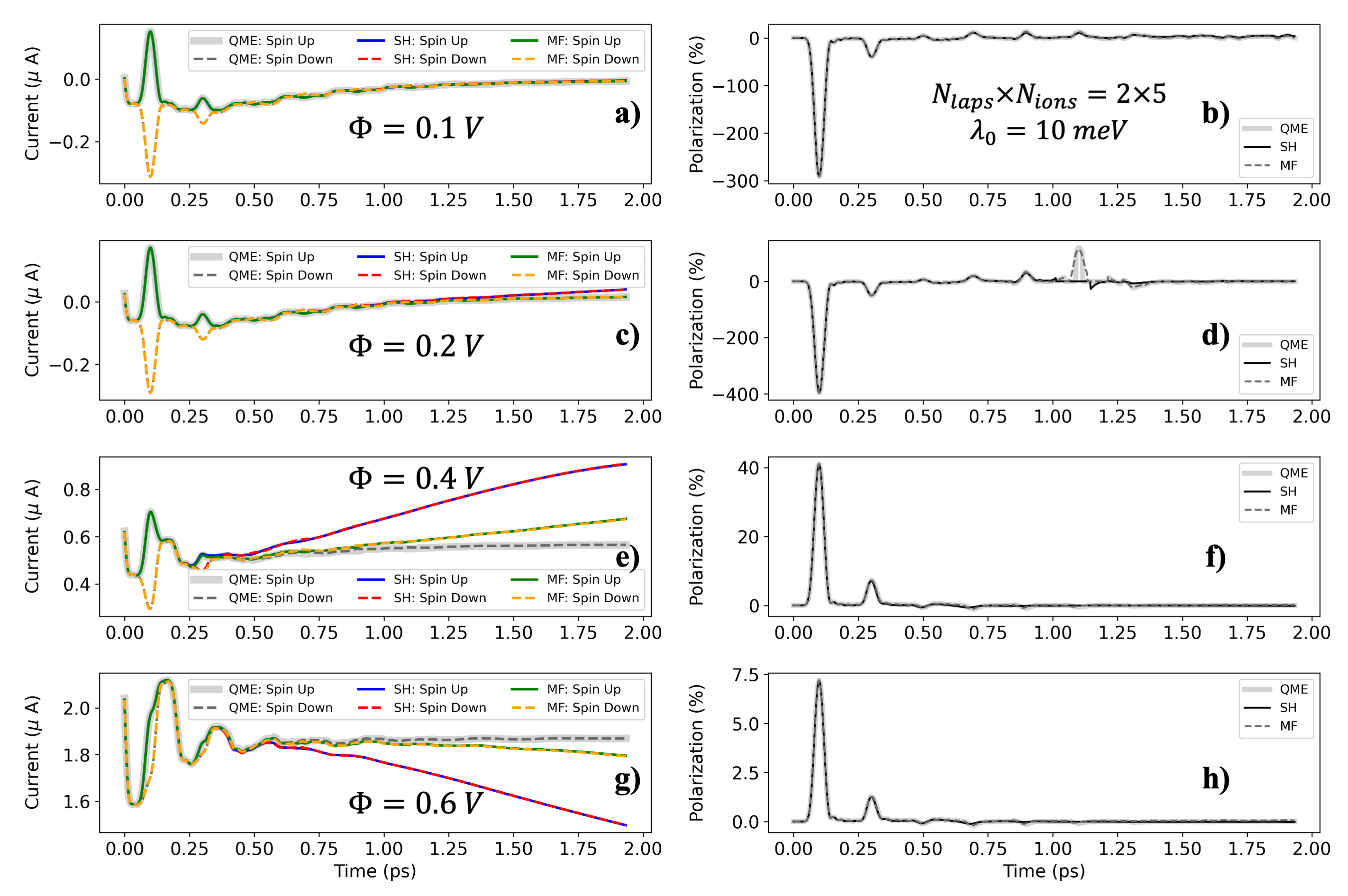}
    \caption{Time-dependent spin current (left panels) and spin polarization (right panels) for a $N_{\rm laps}\times N_{\rm ions}=2\times5$ chiral molecule with electron-phonon couplings under different biases: (a, b) $\Phi=0.1$ V, (c, d) $\Phi=0.2$ V, (e, f) $\Phi=0.4$ V, and (g, h) $\Phi=0.6$ V. Here, we applied a large SOC strength where $\lambda_0=10$ meV.}
    \label{fig:8}
\end{figure}

When the length of the chiral molecule is increased (Fig.~\ref{fig:6}) or the SOC strength is enhanced (Fig.~\ref{fig:7}), the overall bias-dependent trends of the spin current remain qualitatively the same as those observed in the short chiral molecule.
Specifically, electron–phonon coupling continues to play a negligible role at low bias, enhances current through phonon-assisted tunneling at intermediate bias, and suppresses current due to scattering and decoherence at high bias.
For longer molecules, the transient current exhibits stronger amplitude variations and responds more rapidly, showing either enhancement or suppression at earlier times depending on the applied bias voltage (left panels in Fig.~\ref{fig:6}).
With stronger SOC, the ultrafast (fs-scale) variations of the spin current retain the same amplitude as in the case without nuclear motion, while the slower (ps-scale) variations remain unchanged compared to the weak-SOC case for the same molecular length (left panels of Figs.~\ref{fig:7} and~\ref{fig:8}).

Although the ps-scale spin current is influenced by electron-phonon coupling, the spin polarization remains unaffected by nuclear motion across all bias voltages (right panels of Figs.~\ref{fig:6} -~\ref{fig:8}), underscoring that phonon effects alone cannot generate or sustain spin selectivity.

\begin{figure}
    \centering
    \includegraphics[width=0.9\linewidth]{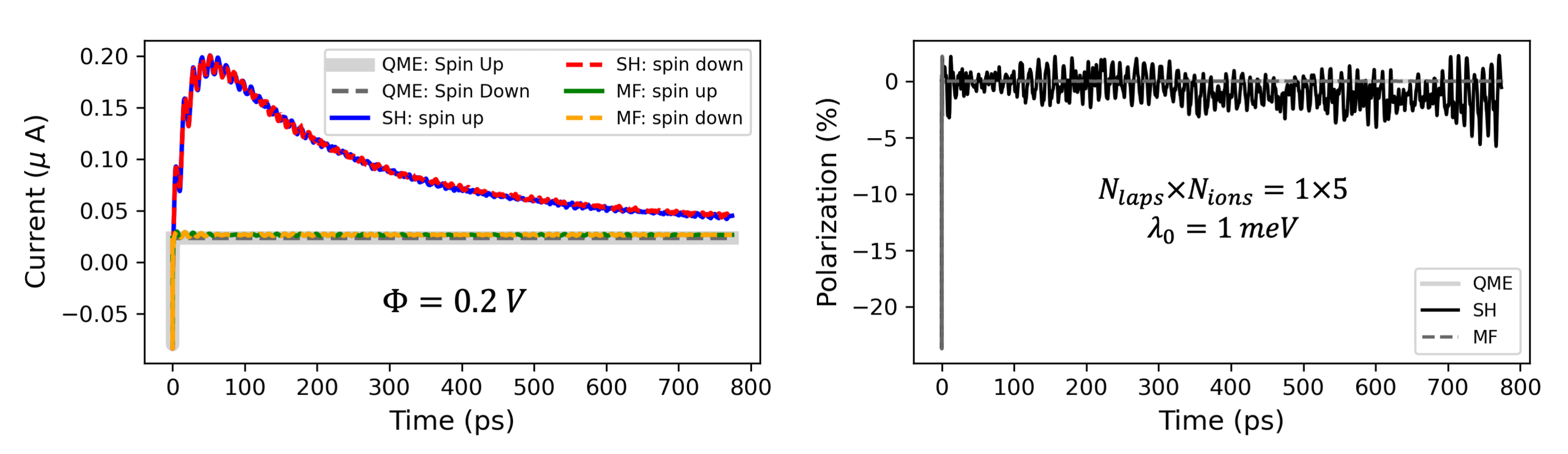}
    \caption{Time-dependent spin current (a) and spin polarization (b) for a $N_{\rm laps}\times N_{\rm ions}=1\times5$ chiral molecule with electron-phonon couplings under a bias of $\Phi=0.2$ V. A small SOC strength of $\lambda_0=1$ meV is applied, and the dynamics are evolved for 800 ps.}
    \label{fig:9}
\end{figure}

Finally, we perform an 800 ps electronic dynamics simulation on a short chiral molecule with $\lambda_0 = 1$ meV and $\Phi = 0.2$ V, as shown in Fig. \ref{fig:9}. Over this long-time evolution, the spin current obtained from SH initially increases and then decreases relative to QME, whereas MF fails to capture this behavior. The overall spin selectivity from SH results fluctuates around zero, indicating that in this regime electron–phonon coupling does not sustain a robust spin polarization. Instead, phonon-induced scattering primarily acts to randomize the spin-dependent transport pathways, preventing the emergence of significant spin selectivity over extended timescales.

\section{THEORETICAL FRAMEWORK}
\subsection{EQUATION OF MOTION}
The general model Hamiltonian in our study consists of three components: the system part (without electron-electron interaction); the bath part; and the system-bath coupling part:
\begin{gather}
    \hat{H}_t=\hat{H}_s+\hat{H}_b+\hat{H}_c, \\
    \hat{H}_s=\sum_{mn}h_{mn}(\bm{R})\hat{d}_m^{\dagger}\hat{d}_n+U(\bm{R})+T(\bm{P}),\\
    \hat{H}_b=\sum_k\epsilon_k\hat{c}_k^{\dagger}\hat{c}_k,\\
    \hat{H}_c=\sum_{kn}V_{kn}(\hat{c}_k^{\dagger}\hat{d}_n+\hat{d}_n^{\dagger}\hat{c}_k)
\end{gather}

It is always possible to transform the one-electron Hamiltonian $\hat{h}$ into the adiabatic representation,
\begin{gather}
    \hat{H}_s=\sum_{mn}h_{mn}(\bm{R})\hat{d}_m^{\dagger}\hat{d}_n+U(\bm{R})+T(\bm{P})\\
    =\sum_{p}\tilde{h}_{pp}(\bm{R})\hat{\tilde{d}}_p^{\dagger}\hat{\tilde{d}}_p+U(\bm{R})+T(\bm{P})
\end{gather}
where $\hat{\tilde{d}}_p=\sum_{m}\Lambda_{pm}\hat{d}_m$, with $\hat{\Lambda}$ being a unitary matrix, such that $\hat{\tilde{h}}$ is diagonal, $\sum_{nm}\Lambda_{pn}^{\dagger}h_{nm}\Lambda_{mq}=\delta_{pq}\tilde{h}_{pq}$. Accordingly, the system-bath coupling Hamiltonian becomes,
\begin{gather}
    \hat{H}_c=\sum_{kp}\tilde{V}_{kp}\left(\hat{c}_k^{\dagger}\hat{\tilde{d}}_p+\hat{\tilde{d}}_p^{\dagger}\hat{c}_k\right), 
\end{gather}
where $\tilde{V}_{kp}=\sum_mV_{km}\Lambda_{mp}$, so the hybridization function $\tilde{\Gamma}_{pq}$ in the wide band approximation is,
\begin{gather}
    \tilde{\Gamma}_{pq}(\bm{R}) =\sum_k\tilde{V}_{kp}^*\tilde{V}_{kq}\delta(\epsilon-\epsilon_k) =\sum_{mn}\Lambda_{pm}^*\Lambda_{nq}\Gamma_{mn}
\end{gather}

We define the single particle reduced density matrix (1-RDM) in the adiabatic representation as,
\begin{gather}
    \tilde{\sigma}_{nm}=Tr_e(\hat{\rho}_{el}\hat{\tilde{d}}_m^{\dagger}\hat{\tilde{d}}_n)
\end{gather}
Starting from the Liouville–von Neumann (LvN) equation\cite{dou2016many,dou2017generalized}, we derive the equation of motion (EOM) for the 1-RDM in the adiabatic representation,
\begin{gather}
    \frac{d}{dt}\hat{\rho}_{el}=-i[\hat{H}_s,\hat{\rho}_{el}]-\hat{\hat{\mathcal{L}}}_{bs}\hat{\rho}_{el},\\
    \frac{d}{dt}\tilde{\sigma}_{nm}=-iTr_e([\hat{H}_s,\hat{\rho}_{el}]\hat{\tilde{d}}_m^{\dagger}\hat{\tilde{d}}_n) -Tr_e(\hat{\hat{\mathcal{L}}}_{bs}\hat{\rho}_{el}\hat{\tilde{d}}_m^{\dagger}\hat{\tilde{d}}_n)
\end{gather}
Following steps in Ref. \citenum{dou2016many}, we can see,
\begin{gather}\label{EOM_adia_1RDM}
    \frac{d}{dt}\hat{\tilde{\sigma}}=-i[\hat{\tilde{h}},\hat{\tilde{\sigma}}]-i[U(\bm{R}),\hat{\tilde{\sigma}}]-i[T(\bm{P}),\hat{\tilde{\sigma}}]-\frac{1}{2}[\hat{\tilde{\Gamma}},\hat{\tilde{\sigma}}]_++\frac{1}{2}[f(\hat{\tilde{h}}),\hat{\tilde{\Gamma}}]_+
\end{gather}
%Accordingly, the EOM for diabatic 1-RDM can be deduced as:
%\begin{gather}
%    \sigma_{nm}=Tr_e(\hat{\rho}_{el}\hat{d}_m^{\dagger}\hat{d}_n), \\
%    \frac{d}{dt}\hat{\sigma}=-i[\hat{h},\hat{\sigma}]-i[U(\bm{R}),\hat{\sigma}]-i[T(\bm{P}),\hat{\sigma}]-\frac{1}{2}[\hat{\Gamma},\hat{\sigma}]_++\frac{1}{2}[Uf(\hat{\tilde{h}})U^{\dagger},\hat{\Gamma}]_+
%\end{gather}
The first three terms in this equation are identical to those in Ref. \citenum{ma2025orbital}, while the last two terms arise from the interaction between the system and the metal bath, where $f(\hat{\tilde{h}})$ is the fermi function.

Without nuclear motion, the QME is
\begin{gather}
    \frac{d}{dt}\hat{\tilde{\sigma}}=-i[\hat{\tilde{h}},\hat{\tilde{\sigma}}]-\frac{1}{2}[\hat{\tilde{\Gamma}},\hat{\tilde{\sigma}}]_++\frac{1}{2}[f(\hat{\tilde{h}}),\hat{\tilde{\Gamma}}]_+
\end{gather}
and the spin current can evaluated by
\begin{equation} \label{eq:current}
\begin{gathered}
    I_L = \mathrm{Tr}\Big(-\frac{1}{2}[\hat{\tilde{\Gamma}}_L, \hat{\tilde{\sigma}}]_+ + \frac{1}{2}[f_L(\hat{\tilde{h}}), \hat{\tilde{\Gamma}}_L]_+\Big),\\
    I_R = \mathrm{Tr}\Big(-\frac{1}{2}[\hat{\tilde{\Gamma}}_R, \hat{\tilde{\sigma}}]_+ + \frac{1}{2}[f_R(\hat{\tilde{h}}), \hat{\tilde{\Gamma}}_R]_+\Big),\\
    I = \frac{1}{2}(I_L + I_R)
\end{gathered}
\end{equation}

\subsection{ORBITAL SURFACE HOPPING ALGORITHM}
Surface hopping (SH) is a mixed quantum-classical approach for simulating molecular dynamics. In this method, nuclear motion is treated classically, with nuclei evolving along Newtonian trajectories and their positions and velocities updated at each time step. Meanwhile, the electronic degrees of freedom are treated quantum mechanically, with the system assumed to evolve on one of several electronic potential energy surfaces (PESs) at any given moment.
Given the EOM for 1-RDM above, we can derive the algorithm of orbital surface hopping for open quantum system.

Firstly, we derive the orbital quantum-classical Liouville equation (QCLE) by performing partial Wigner transformation with respect to the nuclear degree of freedoms (DOFs) on Eq. \ref{EOM_adia_1RDM},
\begin{gather}
    \frac{d}{dt}\hat{\tilde{\sigma}}_W=-i\left((\hat{\tilde{h}}\hat{\tilde{\sigma}})_W-(\hat{\tilde{\sigma}}\hat{\tilde{h}})_W\right)-i\left((U\hat{\tilde{\sigma}})_W-(\hat{\tilde{\sigma}}U)_W\right) - i\left((T\hat{\tilde{\sigma}})_W-(\hat{\tilde{\sigma}}T)_W\right)
    \notag\\ -\frac{1}{2}\left((\hat{\tilde{\Gamma}}\hat{\tilde{\sigma}})_W+(\hat{\tilde{\sigma}}\hat{\tilde{\Gamma}})_W\right)+\frac{1}{2}\left((f(\hat{\tilde{h}})\hat{\tilde{\Gamma}})_W+(\hat{\tilde{\Gamma}}f(\hat{\tilde{h}}))_W\right),
\end{gather}
Note that the Wigner-Moyal
operator can be used to express the partial Wigner transform
of the product of operators $\hat{A}$ and $\hat{B}$:
\begin{gather}
    (\hat{A}\hat{B})_W=\hat{A}_We^{-i\overleftrightarrow{\Lambda}/2}\hat{B}_W,\\
    \overleftrightarrow{\Lambda} = \overleftarrow{\frac{\partial}{\partial P}}\overrightarrow{\frac{\partial}{\partial R}}-\overleftarrow{\frac{\partial}{\partial R}}\overrightarrow{\frac{\partial}{\partial P}}.
\end{gather}
When truncating the Wigner-Moyal operator to the first order
in the Taylor expansion, $e^{-i\overleftrightarrow{\Lambda}/2} \approx 1- i\hbar\overleftrightarrow{\Lambda}/2$, we arrived at orbital QCLE,
\begin{gather}
    \frac{d}{dt}\hat{\tilde{\sigma}}_W=-i[\hat{\tilde{h}}_W,\hat{\tilde{\sigma}}_W]-\frac{P}{M}\frac{\partial\hat{\tilde{\sigma}}_W}{\partial R} + \frac{\partial U}{\partial R}\frac{\partial\hat{\tilde{\sigma}}_W}{\partial P}+\frac{1}{2}\left[\frac{\partial\hat{\tilde{h}}_W}{\partial R},\frac{\partial\hat{\tilde{\sigma}}_W}{\partial P}\right]_+ \notag\\
    -\frac{1}{2}[\hat{\tilde{\Gamma}}_W,\hat{\tilde{\sigma}}_W]_++\frac{1}{2}[f(\hat{\tilde{h}}_W),\hat{\tilde{\Gamma}}_W]_+
\end{gather}
Note that we only keep the zeroth-order gradient
expansion for the last two terms, following Ref. \citenum{dou2017generalized}.
In the orbital surface hopping, we propagate the orbital density matrix for each trajectory according to
\begin{gather}\label{EOM_sigma}
    \frac{d\tilde{\sigma}_{nm}}{dt}=- \frac{P}{M}\sum_l(\mathcal{D}_{nl}\tilde{\sigma}_{lm}-\tilde{\sigma}_{nl}\mathcal{D}_{lm})-\frac{1}{2}\sum_l(\tilde{\Gamma}_{nl}\tilde{\sigma}_{lm}+\tilde{\sigma}_{nl}\tilde{\Gamma}_{lm}) + \frac{1}{2}\left(f(\tilde{h}_{nn})\tilde{\Gamma}_{nm}+\tilde{\Gamma}_{nm}f(\tilde{h}_{mm})\right),
\end{gather}
where $\mathcal{D}_{nm}$ is the derivative coupling 
\begin{gather}
    \mathcal{D}_{nm}=\sum_k\Lambda_{kn}^*\frac{\partial \Lambda_{km}}{\partial R},
\end{gather}
For the EOM of nucleus,
\begin{gather}
    \dot{\bm{R}}=\frac{\bm{P}}{M}, \\
    \dot{\bm{P}}=-\frac{\partial U}{\partial\bm{R}} +\sum_{i=1}^nF_{\lambda_i,\lambda_i},
\end{gather}
where $F_{\lambda_i,\lambda_i}$ is the force of the occupied surface $\lambda_i$.

There are two types of hopping rates: those arising from the derivative coupling within the system, denoted as $k_{n\rightarrow m}^d$, and those arising from electron transfer between the system and the bath, denoted as $k^L$.
For $k_{n\rightarrow m}^d$, we have
\begin{gather}
    k_{n\rightarrow m}^d=max\left[2\mathcal{R}\frac{P}{M}\frac{\mathcal{D}_{nm}\tilde{\sigma}_{mn}}{\tilde{\sigma}_{nn}},0\right]
\end{gather}
For the hopping rate originating from the metal surface, we only consider inter-state hopping, since the electrons on the metal surface orbitals are traced out.
This implies that there is no direct hopping between the system orbitals and the bath orbitals. Consequently, each system orbital is treated as having only two possible states: occupied (denoted as 1) or unoccupied (denoted as 0).
This issue has been previously resolved\cite{dou2015surface,dou2015surface3}:
\begin{gather}
    k_{nn(0\rightarrow 1)}^L=\tilde{\Gamma}_{nn}f(\tilde{h}_{nn}),\\
    k_{nn(1\rightarrow 0)}^L=\tilde{\Gamma}_{nn}\left(1-f(\tilde{h}_{nn})\right)
\end{gather}
The spin current can be evaluated from Eq.~\ref{eq:current}.

\begin{figure}
    \centering
    \includegraphics[width=0.7\linewidth]{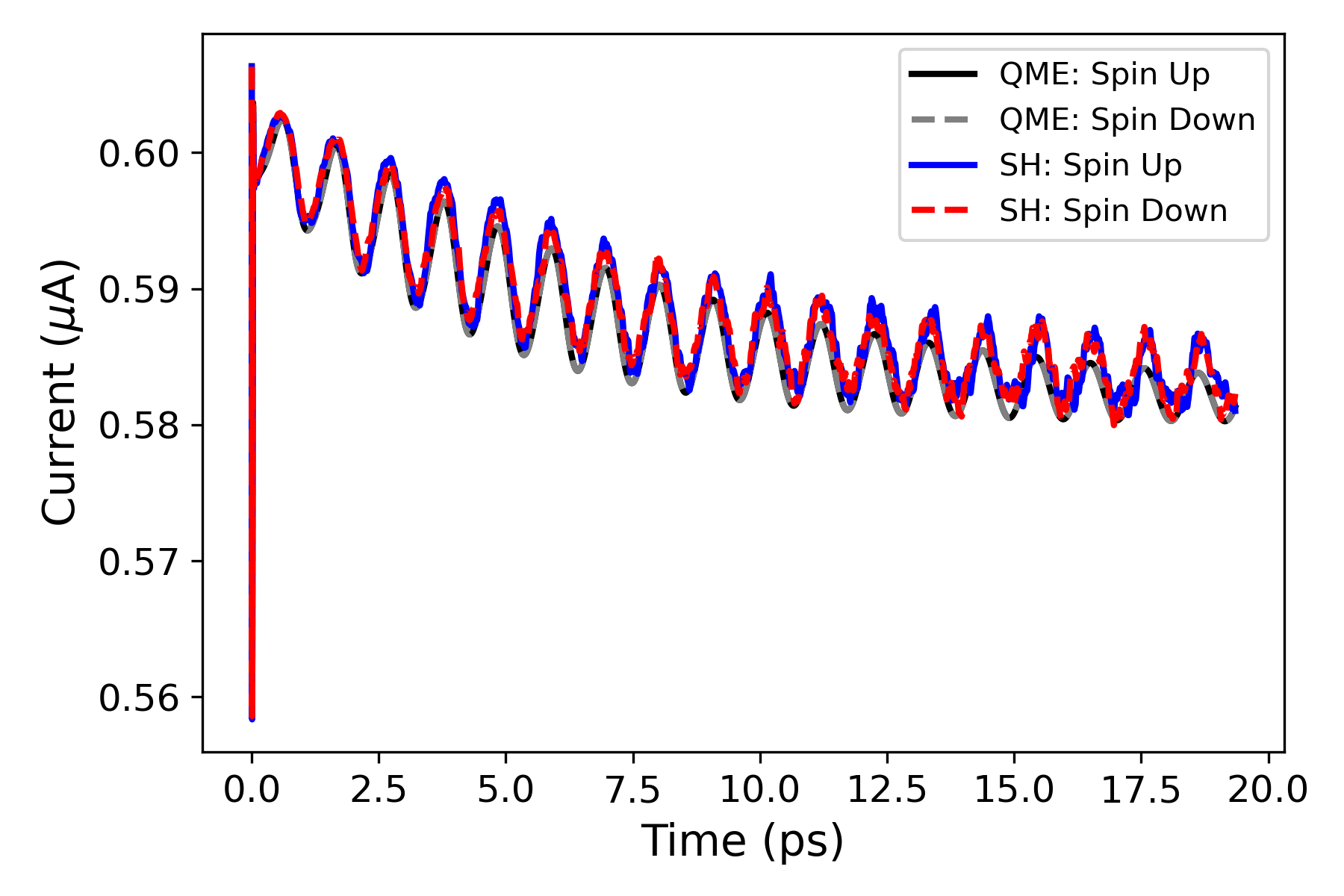}
    \caption{Time-dependent spin current for both QME and SH approaches.}
    \label{fig:10}
\end{figure}

We benchmarked this SH algorithm with the QME using a simple 2-level model for the system as:
\begin{gather}
    \hat{H}_s=
    \begin{bmatrix}
        \epsilon_0 & -(t_0+it_0)+(t_1+it_1)(a^{\dagger}+a) \\
        -(t_0-it_0)+(t_1-it_1)(a^{\dagger}+a) & \epsilon_0
    \end{bmatrix}
    + \hbar\omega_0(a^{\dagger}a+\frac{1}{2})
\end{gather}
where $\epsilon_0=-240$ meV, $t_0=40$ meV, $t_1=4$ meV, $\hbar\omega_0=4$ meV, while all other parameters are the same as those in Eq.~\ref{H_chiral}. 
The quantized vibrational mode is described by bosonic creation and annihilation operators $a^{\dagger}$ and $a$.
We plot the spin current dynamics over 20 ps results both from QME (150 phonon basis) and SH (averaged over 10000 trajectories) in Fig.~\ref{fig:10}.
The results show excellent agreement between the two methods, confirming the reliability of the approach presented here.

\subsection{ORBITAL MEAN FIELD EHRENFEST ALGORITHM}

The mean-field Ehrenfest (MF) method is a mixed quantum–classical approach in which the electronic degrees of freedom are treated quantum mechanically, while the nuclei evolve classically on an averaged potential energy surface determined by the instantaneous electronic state.
Therefore, the EOM of electronic density matrix is the same with that in SH algorithm (Eq. \ref{EOM_sigma}), while the EOM of nuclear part becomes much easier than SH:
\begin{gather}
    \dot{\bm{R}}=\frac{\bm{P}}{M}, \\
    \dot{\bm{P}}=Tr\left(\frac{\partial\hat{H}_s}{\partial\bm{R}}\hat{\sigma}\right)
\end{gather}

\section{CONCLUSIONS}

In this work, we systematically investigated spin transport in chiral molecular junctions with and without electron–phonon coupling. Using the QME for purely electronic dynamics and SH together with MF approaches for electron–phonon interactions, we characterized the time-dependent behavior of spin current and spin polarization under varying bias voltage, molecular length, and SOC strength.

Our results demonstrate that transient spin polarization emerges within the first few hundred femtoseconds but decays to zero at long times, independent of electron–phonon coupling. Higher bias voltages increase the overall spin current but suppress spin polarization, while longer molecules and stronger SOC enhance transient polarization. Including electron–phonon coupling further modifies current–voltage characteristics, leading to current enhancement at intermediate bias but suppression at high bias, while leaving polarization dynamics largely unaffected.
The SH method incorporates stochastic transitions between adiabatic potential energy surfaces, thereby providing a more accurate representation of phonon-induced broadening in transport channels, whereas the MF approach, due to its averaged mean-field character, systematically underestimates this effect.

These findings clarify the interplay of molecular length, SOC, and vibrational effects in CISS, and they emphasize the importance of considering both electronic structure and vibrational effects when designing molecular spintronic systems. Looking forward, the approaches developed here may be extended to more complex junction architectures or combined with cavity quantum electrodynamics to explore light–matter–spin interactions in chiral systems.

\begin{acknowledgement}
W.D. thanks the funding from National Natural Science Foundation of China (No. 22361142829) and Zhejiang Provincial Natural Science Foundation (No. XHD24B0301). Y.W. acknowledges supports from the National Natural Science Foundation of China No. 22403077.

\end{acknowledgement}

%\bibliography{rsc}
\bibliography{acs}

\end{document}